\documentclass[preprint,amsmath,amssymb,aps,prd,showpacs,10pt]{revtex4-1}

\usepackage{amsmath,amsfonts,amssymb}
\usepackage{hyperref,bbm}

\newcommand{\ud}{\mathrm{d}}

\newcommand{\om}{\omega_{\textrm{\tiny BD}}}

\begin{document}

\title{Brans-Dicke theory and the emergence of \boldmath{$\Lambda$}CDM model}

\author{Orest Hrycyna}
\email{orest.hrycyna@fuw.edu.pl}
\affiliation{Theoretical Physics Division, National Centre for Nuclear
Research,  Ho{\.z}a 69, 00-681 Warszawa, Poland}
\author{Marek Szyd{\l}owski}
\email{marek.szydlowski@uj.edu.pl}
\affiliation{Astronomical Observatory, Jagiellonian University, Orla 171,
30-244 Krak{\'o}w, Poland}
\affiliation{Mark Kac Complex Systems Research Centre, Jagiellonian
University, Reymonta 4, 30-059 Krak{\'o}w, Poland}

\date{\today}

\begin{abstract}
The dynamics of the Brans-Dicke theory with a scalar field potential function is investigated.
We show that the system with a barotropic matter content can be reduced to an
autonomous three-dimensional dynamical system. For an arbitrary potential function we found the 
values of the Brans-Dicke
parameter for which a global attractor in the phase space representing de Sitter state exists. 
Using linearized solutions in the vicinity of this
critical point we show that the evolution of the Universe mimics the $\Lambda$CDM
model. From the recent Planck satellite data, we obtain constraints on the variability of the 
effective gravitational coupling constant as well as the lower limit of the mass of the Brans-Dicke scalar 
field at the de Sitter state.
\end{abstract}

\maketitle

\section{Introduction}
 
Recent astronomical observations \cite{Ade:2013lta} indicate that the standard cosmological 
model 
($\Lambda$CDM) is still a good effective theory of the physical Universe, although some 
anomalies 
in the power spectrum of CMB are predicted \cite{Ade:2013sta}.

In this theory the cosmological parameters, which are estimated from the 
astronomical observations, play a crucial role. The nature of some parameters, like density parameters for the dark 
sector 
of the Universe (dark matter and dark energy), is still unknown and we are looking 
for 
an explanation. We believe that a more fundamental theory of gravity will reveal the nature of 
the cosmological parameters. In other words we expect that the nature of these parameters is 
emergent.

The Brans-Dicke theory of gravity \cite{Brans:1961sx} seems to be an interesting way to move toward 
construction of a more
adequate description of the evolution of the Universe that includes an explanation of why the Universe enters the accelerating phase of expansion only in the current epoch.

In this paper, we show how dynamical evolution of the standard cosmological
model emerges from the 
cosmological model in which we use the Brans-Dicke theory of 
gravity instead of general relativity. In principle, we obtain two different evolutionary scenarios in which the $\Lambda$CDM 
model with some additional corrections is recovered. Those new additive terms appear in the 
equation describing the evolution of the Universe and the parameters describing the dark 
sector are modified. The generalized Friedmann-Robertson-Walker equation can be tested through 
the 
astronomical observations, and the model itself can be falsified be the data.

The gravitational sector of the Brans-Dicke theory is modified by the presence of some potential 
function, while the barotropic matter is used for description of the matter content of the 
Universe. However, because of the ambiguity of the form of the potential (for some new 
results concerning the scalar field potential function for the inflaton field see recent Planck 
results \cite{Ade:2013rta}), our strategy is to obtain results as soon as possible without the ``potential bias''. Therefore, in our analysis we concentrate on the behavior of the dynamical 
system describing the evolution of the Universe in the vicinity of a special critical point 
whose position in the phase space does not depend on the detailed form of the potential 
function. 

We demonstrate that there are two different types of emergence of the $\Lambda$CDM model. 
Different initial conditions and the model parameters give rise to different scenarios. In the 
first scenario, the $\Lambda$CDM model is approached monotonically, while in the second scenario 
we obtain an oscillatory approach to the de Sitter state. In both scenarios, the leading term 
represents 
evolution in the $\Lambda$CDM model. 

Only the astronomical data can help in selecting the type of evolution that is realized by our 
Universe. Alternatively, deeper knowledge of the initial conditions and the model parameters
can distinguish a single type of the behavior.

\section{The Brans-Dicke cosmology}

The action for the Brans-Dicke theory \cite{Brans:1961sx} in the so-called Jordan
frame is in the following form \cite{Faraoni:book,Capozziello:book},
\begin{equation}
\label{eq:action}
S = \int\ud^{4}x\sqrt{-g}\left\{\phi R
-\frac{\om}{\phi}\nabla^{\alpha}\phi\nabla_{\alpha}\phi - 
2\,V(\phi)\right\} + 16\pi S_m\,,
\end{equation}
where the barotropic matter is described by
\begin{equation}
S_{m} = \int\ud^{4}x\sqrt{-g}\mathcal{L}_{m}\,,
\end{equation}
and $\om$ is a dimensionless parameter of the theory.

Variation of the total action (\ref{eq:action}) with respect to the metric
tensor $\delta S/\delta g^{\mu\nu} =0$ gives the field equations for the theory
\begin{equation}
\label{eq:feq}
\phi\left(R_{\mu\nu}-\frac{1}{2}g_{\mu\nu}R\right)
 =\frac{\om}{\phi}\left(\nabla_{\mu}\phi\,\nabla_{\nu}\phi -
\frac{1}{2}\,g_{\mu\nu}\,\nabla^{\alpha}\phi\,\nabla_{\alpha}\phi\right)
- g_{\mu\nu}V(\phi)
-\left(g_{\mu\nu}\Box\phi-\nabla_{\mu}\nabla_{\nu}\phi\right) + 8\pi\,
T_{\mu\nu}^{(m)}\,,
\end{equation}
where the energy momentum tensor for the matter content is 
\begin{equation}
T_{\mu\nu}^{(m)} = -\frac{2}{\sqrt{-g}}\,\frac{\delta}{\delta
g^{\mu\nu}}\left(\sqrt{-g}\mathcal{L}_{m}\right)\,.
\end{equation}
Taking the trace of \eqref{eq:feq} one obtains
\begin{equation}
\label{eq:trace}
R = \frac{\om}{\phi^{2}}\,\nabla^{\alpha}\phi\,\nabla_{\alpha}\phi +
4\frac{V(\phi)}{\phi} + 3 \frac{\Box\phi}{\phi} - \frac{8\pi}{\phi} T^{m}.
\end{equation}
Variation of the action with respect to $\phi$ gives
\begin{equation}
\Box\phi = \frac{1}{2\phi}\nabla^{\alpha}\phi\,\nabla_{\alpha}\phi -
\frac{\phi}{2\om}\left(R - 2 V'(\phi)\right)\,,
\end{equation}
and using \eqref{eq:trace} to eliminate the Ricci scalar $R$, one obtains
\begin{equation}
\label{eq:sf_mo}
\Box\phi = -\frac{2}{3+2\om}\left(2 V(\phi) - \phi V'(\phi)\right) +
\frac{8\pi}{3+2\om} T^{m}\,.
\end{equation}

The field equations \eqref{eq:feq} are adopted to the spatially flat Friedmann-Robertson-Walker 
metric
\begin{equation}
\ud s^{2} = -\ud t^{2} + a^{2}(t)\big(\ud x^{2} + \ud y^{2} + \ud z^{2}\big)\,
\end{equation}
and the matter content is described by the barotropic equation of state $p_{m}
= w_{m}\rho_{m}$, where $p_{m}$ and $\rho_{m}$ are the pressure and the energy
density of the matter.

The energy conservation condition is
\begin{equation}
\label{eq:encon:1}
3 H^{2} = \frac{\om}{2}\frac{\dot{\phi}^{2}}{\phi^{2}} + \frac{V(\phi)}{\phi} -
3 H \frac{\dot{\phi}}{\phi} + \frac{8\pi}{\phi}\rho_{m}
\end{equation}
and the acceleration equation is
\begin{equation}
\label{eq:acc:1}
\dot{H} =
-\frac{\om}{2}\frac{\dot{\phi}^{2}}{\phi^{2}}-\frac{1}{3+2\om}\frac{2\,V(\phi)-\phi
V'(\phi)}{\phi} + 2 H \frac{\dot{\phi}}{\phi} -
\frac{8\pi}{\phi}\rho_{m}\frac{2+\om(1+w_{m})}{3+2\om}\,.
\end{equation}
The dynamical equation \eqref{eq:sf_mo} for the scalar field reduces to
\begin{equation}
\ddot{\phi}+3H\dot{\phi}=2\frac{2\,V(\phi)-\phi\,V'(\phi)}{3+2\om} +
8\pi\rho_{m}\frac{1-3w_{m}}{3+2\om}\,.
\end{equation}

In what follows we introduce the following energy phase space variables
\begin{equation}
x\equiv \frac{\dot{\phi}}{H \phi}\,,\qquad y\equiv
\sqrt{\frac{V(\phi)}{3\phi}}\frac{1}{H}\,,\qquad
\lambda\equiv-\phi\frac{V'(\phi)}{V(\phi)} \,.
\end{equation}
Then the energy conservation condition (\ref{eq:encon:1}) can be presented as
\begin{equation}
\frac{8\pi \rho_{m}}{3 H^{2} \phi} = 1 + x -\frac{\om}{6}x^{2} - y^{2}
\label{eq:concond}
\end{equation}
and the acceleration equation (\ref{eq:acc:1})
\begin{equation}
\label{eq:acc:2}
	\frac{\dot{H}}{H^{2}} = 2x-\frac{\om}{2}x^{2}
	-\frac{3}{3+2\om}y^{2}(2+\lambda)
-
3\left(1 + x -\frac{\om}{6}x^{2} - y^{2}\right)\frac{2+\om(1+w_{m})}{3+2\om}\,.
\end{equation}

The dynamical system describing the evolution of the Brans-Dicke theory of
gravity with a scalar field potential and the barotropic matter has the
following form,
\begin{align}
\label{eq:dynsys}
\frac{\ud x}{\ud \tau} & =  -3x - x^{2} - x \frac{\dot{H}}{H^{2}} +
\frac{6}{3+2\om}y^{2}(2+\lambda) +
3\left(1 + x -\frac{\om}{6}x^{2} - y^{2}\right)\frac{1-3w_{m}}{3+2\om}\,,
\nonumber \\
\frac{\ud y}{\ud \tau} & =  -y\left(\frac{1}{2}x(1+\lambda)+\frac{\dot{H}}{H^{2}}\right)\,, \\
\frac{\ud \lambda}{\ud \tau} & =  x\lambda\Big(1-\lambda(\Gamma-1)\Big)\,,\nonumber
\end{align}
where $\frac{\ud}{\ud \tau}=\frac{\ud}{\ud \ln{a}}$ and
\begin{equation}
\Gamma=\frac{V''(\phi)V(\phi)}{V'(\phi)^{2}}\,,
\end{equation}
where $()'=\frac{\ud}{\ud\phi}.$

From now on we will assume that $\Gamma=\Gamma(\lambda)$. The critical points of
the system (\ref{eq:dynsys}) depend on the explicit form of the
$\Gamma(\lambda)$ function. One can notice that the critical points $(x^{*}=0\,,
y^{*}=\pm1\,,\lambda^{*}=-2)$ do not depend on the assumed $\Gamma(\lambda)$.
Additionally, the acceleration (\ref{eq:acc:2}) calculated at these
points vanishes, giving rise to the de Sitter evolution.

Our analysis concentrates on the critical point ($x^{*}=0$, $y^{*}=1$, $\lambda^{*}=-2$) which 
corresponds to the de Sitter state, while the critical point with $y^{*}=-1$ 
corresponds to the anti--de Sitter state.

The linearization matrix calculated at this point is
\begin{equation}
A=\left(
\begin{array}{ccc}
-3\frac{2+2 \om+3 w_{m}}{3+2\om} & -6\frac{1-3 w_{m}}{3+2\om} & \frac{6}{3+2\om} \\
     \frac{3}{2}\frac{1+2 \om w_{m}}{3+2\om} & -6\frac{2+\om(1+w_{m})}{3+2\om} &
     \frac{3}{3+2\om} \\
     \frac{3}{8}\delta & 0 & 0
\end{array}\right)\,,
\end{equation}
where we assumed $\frac{\partial\lambda'}{\partial
x}\Big|_{*}=\lambda^{*}\Big(1-\lambda^{*}(\Gamma(\lambda^{*})-1)\Big)=\frac{3}{8}\delta$,
and $\delta$ is a nonzero constant and $\frac{\partial\Gamma(\lambda)}{\partial\lambda}\Big|_{*}$ 
is finite. The eigenvalues are
\begin{equation}
l_{1}=-3(1+w_{m})\,, \quad
l_{2,3}=-\frac{3}{2}\left(1\pm\sqrt{\frac{3+2\om+\delta}{3+2\om}}\right)\,.
\end{equation}
For $\delta=0$ one of the eigenvalues vanishes and this critical point is
degenerated. Therefore, the analysis based on the linearization matrix is not
enough and the center manifold theorem must be used \cite{Perko:2001, 
Wiggins:2003, Hrycyna:2010yv}.

The critical point under consideration is stable when $w_{m}>-1$ and
\begin{equation}
\frac{\delta}{3+2\om}<0\,.
\end{equation}
In this case the critical point is a stable node for $-1<\frac{\delta}{3+2\om}<0$ or a stable focus for $\frac{\delta}{3+2\om}<-1$. This indicates that for a given family of potential functions described by the 
function $\Gamma(\lambda)$, we can always find ranges of the $\om$ parameter for which the 
critical point under consideration is a stable one. Let us consider two simple potential 
functions. For potentials of the type $V(\phi)= V_{0}(\phi^{2}-v^{2})^{2}$, we have $\delta=-
\frac{16}{3}$, and for $\om>\frac{7}{6}$ we have a stable node critical point, while for $-
\frac{3}{2}<\om<\frac{7}{6}$ we have a stable focus. It represents a saddle type critical point 
otherwise. For the potential type $V(\phi)=\frac{1}{2}m^{2}\phi^{2}+\frac{\alpha}{4}\phi^{4}$, 
we have $\delta=\frac{8}{3}$. For $\om<-\frac{17}{6}$ this critical point represents a stable 
node, and for $-\frac{17}{6}<\om<-\frac{3}{2}$ we have the stable focus. 

We need to consider both types of behavior separately. 

In the first case we use the following substitution,
\begin{equation}
\frac{\delta}{3+2\om} = \frac{4}{9}n(n-3)\,,
\end{equation}
where for $0<n<\frac{3}{2}$ we have a stable node critical point. The eigenvalues of the 
linearization matrix are
\begin{equation}
l_{1}=-3(1+w_{m})\,, \quad l_{2}=-n\,, \quad l_{3}=-3+n\,,
\end{equation}
and the linearized solutions are
\begin{subequations}
\begin{eqnarray}
x(\tau) & = & 4\frac{n(n-3)(1+w_{m})(1-3w_{m})}{\delta(n-3(1+w_{m}))(n+3w_{m})}(\Delta x-2\Delta 
y) \exp{(-3(1+w_{m})\tau)}+ \nonumber \\ & & + \frac{n}{3\delta(2n-3)(n-3(1+w_{m}))}
\bigg( -4n(n-3)(1-3w_{m})(\Delta x - 2\Delta y) 
+ \nonumber \\ & & \hspace{5cm} 
+(n-3(1+w_{m}))
(3\delta\Delta x -8(n-3)\Delta\lambda)\bigg)\exp{(-n\tau)} + \nonumber \\ & &
+ \frac{n-3}{3\delta(2n-3)(n+3w_{m})}\bigg( 4n(n-3)(1-3w_{m})(\Delta x - 2\Delta y) + 
\nonumber \\ & & \hspace{4cm} + 
(n+3w_{m})(3\delta\Delta x +8n\Delta \lambda)\bigg) \exp{\big((-3+n)\tau
\big)}\,,\\
\label{eq:linT_1a}
y(\tau) & = & 1 + \bigg(2\frac{n(n-3)(1+w_{m})(1-3w_{m})}{\delta(n-3(1+w_{m}))(n+3w_{m})} - 
\frac{1}{2}\bigg)\big(\Delta x -2 \Delta y\big)\exp{\big(-3(1+w_{m})\tau\big)} + \nonumber \\ & 
& + \frac{n}{6\delta(2n-3)(n-3(1+w_{m}))}\bigg(-4n(n-3)(1-3w_{m})(\Delta x -2 \Delta y)
+ \nonumber \\ & & \hspace{5cm} 
+ (n-3(1+w_{m}))(3\delta\Delta x - 8(n-3)\Delta \lambda)\bigg)
\exp{\big(-n\tau\big)} + \nonumber \\ & &
+ \frac{n-3}{6\delta(2n-3)(n+3w_{m})}\bigg(4n(n-3)(1-3w_{m})(\Delta x - 2\Delta y) 
+ \nonumber \\ & & \hspace{4cm} 
+ (n+3w_{m})(3\delta\Delta x + 8n \Delta\lambda)\bigg)\exp{\big((-3+n)\tau
\big)}\,,\\
\lambda(\tau) & = & -2 -
\frac{n(n-3)(1-3w_{m})}{2(n-3(1+w_{m}))(n+3w_{m})}(\Delta x-2\Delta y)\exp{\big(-3(1+w_{m})\tau
\big)} + \nonumber \\ & &
+\frac{1}{8(2n-3)(n-3(1+w_{m}))}\bigg(4n(n-3)(1-3w_{m})(\Delta x - 2\Delta y) 
- \nonumber \\ & & \hspace{5cm} 
- (n-3(1+w_{m}))(3\delta\Delta x - 8(n-3)\Delta\lambda)\bigg)\exp{\big(-n\tau\big)} 
+ \nonumber \\ & &
+ \frac{1}{8(2n-3)(n+3w_{m})}\bigg(4n(n-3)(1-3w_{m})(\Delta x -2\Delta y)
+ \nonumber \\ & & \hspace{4cm} 
+(n+3w_{m})(3\delta\Delta x + 8n\Delta\lambda)\bigg)\exp{\big((-3+n)\tau\big)}\,
\end{eqnarray}
\end{subequations}
where $\Delta x = x^{(i)}$, $\Delta y = y^{(i)}-1$, and $\Delta\lambda= \lambda^{(i)}+2$ are the 
initial conditions.

Using these linearized solutions and the conservation condition \eqref{eq:concond}, we can find 
a linearized solution for the barotropic matter density fraction. Up to linear terms in initial 
conditions, we obtain
\begin{equation}
\label{eq:Omega_m1}
\Omega_{m} \approx (\Delta x - 2 \Delta y)\exp{\big(-3(1+w_{m})\tau\big)} = \Omega_{m,i}
\exp{\big(-3(1+w_{m})\tau\big)}.
\end{equation}

Now, using the linearized solutions, we are ready to compute the Hubble function. The 
acceleration equation \eqref{eq:acc:2} can be cast into the following form,
\begin{equation}
\label{eq:acc:3}
\frac{\ud\ln{H^{2}}}{\ud\tau} = 2 \frac{\dot{H}}{H^{2}} = 4x-\om x^{2}
	-\frac{6}{3+2\om}y^{2}(2+\lambda)
-6\left(1 + x -\frac{\om}{6}x^{2} - y^{2}\right)\frac{2+\om(1+w_{m})}{3+2\om}\,,
\end{equation}
after integration of this equation, and up to the linear terms in initial conditions, we obtain
\begin{equation}
\left(\frac{H(a)}{H(a_{0})}\right)^{2} \approx \Omega_{\Lambda,0} +
\Omega_{M,0}\left(\frac{a}{a_{0}}\right)^{-3(1+w_m)} +
\Omega_{n,0}\left(\frac{a}{a_{0}}\right)^{-n} +
\Omega_{3n,0}\left(\frac{a}{a_{0}}\right)^{-3+n} \,,
\label{eq:H2T_1}
\end{equation}
where
\begin{subequations}
\begin{align}
\Omega_{M,0}  = &\bigg(1- 
\frac{4n(n-3)(1-3w_{m})(4+3w_{m})}{3\delta(n+3w_{m})(n-3(1+w_{m}))}\bigg)\,\Omega_{m,0}\,,\\
\Omega_{n,0} =  &
\frac{n+1}{3\delta(2n-3)}\bigg(\frac{4n(n-3)(1-3w_{m})}{n-3(1+w_{m})}\,\Omega_{m,i} -3\delta\,
\Delta x+8(n-3)\Delta\lambda\bigg)\left(\frac{a_{0}}{a^{(i)}}\right)^{-n}\,,\\
\Omega_{3n,0} = & 
\frac{n-4}{3\delta(2n-3)}\bigg(-\frac{4n(n-3)(1-3w_{m})}{n+3w_{m}}\,\Omega_{m,i}
-3\delta\,\Delta x -8n\Delta\lambda\bigg) \left(\frac{a_{0}}{a^{(i)}}\right)^{-3+n} \,,
\end{align}
\label{eq:Omega_1}
\end{subequations}
and
\begin{equation}
\Omega_{\Lambda,0} = 1- \Omega_{M,0}-\Omega_{n,0}-\Omega_{3n,0}\,,
\end{equation}
and we have used that in the linear approximation
\begin{equation}
\Omega_{m,0} = \Omega_{m,i}\left(\frac{a_{0}}{a^{(i)}}\right)^{-3(1+w_{m})}\,.
\end{equation}

Now we proceed to the investigation of the stable focus critical point. In this case we 
use the substitution
\begin{equation}
\frac{\delta}{3+2\om}=-\frac{1}{9}(9+4n^{2})\,,
\end{equation}
and the eigenvalues of the linearization matrix are
\begin{equation}
l_{1}=-3(1+w_{m})\,, \quad l_{2}=-\frac{3}{2} - \mathbbmtt{i} n\,, \quad l_{3}=-\frac{3}{2} + 
\mathbbmtt{i} n\,. 
\end{equation}
The linearized solutions of the system are
\begin{subequations}
\begin{eqnarray}
x(\tau) & = & \frac{4(4n^{2}+9)(1+w_{m})(1-3w_{m})}{\delta(4n^{2}+9(1+2w_{m})^{2})} 
\big(\Delta x - 2\Delta y\big) \exp{\big(-3(1+w_{m})\tau\big)} + \nonumber \\ & &
+\bigg(-\frac{4(4n^{2}+9)(1+w_{m})(1-3w_{m})}{\delta(4n^{2}+9(1+2w_{m})^{2})}\big(\Delta 
x-2\Delta y\big) +\Delta x\bigg)\exp{\big(-\frac{3}{2}\tau\big)}\cos{\big(n\tau\big)} + 
\nonumber \\
& & +\frac{1}{6n}\bigg(-\frac{2(4n^{2}+9)(4n^{2}-9(1+2w_{m}))(1-3w_{m})}
{\delta(4n^{2}+9(1+2w_{m})^{2})}\big(\Delta x-2\Delta y\big)
- \nonumber \\& & \hspace{1.25cm} 
- 9\Delta x-\frac{4}{\delta}(4n^{2}+9)\Delta\lambda\bigg)\exp{\big(-\frac{3}
{2}\tau\big)}\sin{\big(n\tau\big)}\,,\\
y(\tau) & = & 1+\frac{1}{2}\bigg(\frac{4(4n^{2}+9)(1+w_{m})(1-3w_{m})}
{\delta(4n^{2}+9(1+2w_{m})^{2})}-1\bigg)\big(\Delta x -2 \Delta y\big)\exp{\big(-3(1+w_{m})\tau
\big)}+ \nonumber \\ & &
+\frac{1}{2}\bigg(-\frac{4(4n^{2}+9)(1+w_{m})(1-3w_{m})}{\delta(4n^{2}+9(1+2w_{m})^{2})} 
\big(\Delta x -2\Delta y\big)+\Delta x\bigg)\exp{\big(-\frac{3}{2}\tau\big)}\cos{\big(n\tau
\big)}
+ \nonumber \\ & & 
+\frac{1}{12n}\bigg(-\frac{2(4n^{2}+9)(4n^{2}-9(1+2w_{m}))(1-3w_{m})}
{\delta(4n^{2}+9(1+2w_{m})^{2})}\big(\Delta x-2\Delta y\big)
- \nonumber \\& & \hspace{1.25cm} 
- 9\Delta x-\frac{4}{\delta}(4n^{2}+9)\Delta\lambda\bigg)\exp{\big(-\frac{3}
{2}\tau\big)}\sin{\big(n\tau\big)}\,,\\
\lambda(\tau) & =& -2-\frac{(4n^{2}+9)(1-3w_{m})}{2(4n^{2}+9(1+2w_{m})^{2})} 
\big(\Delta x -2 \Delta y\big)\exp{\big(-3(1+w_{m})\tau\big)} + \nonumber \\ 
& & + \bigg(\frac{(4n^{2}+9)(1-3w_{m})}{2(4n^{2}+9(1+2w_{m})^{2})}\big(\Delta x -2 \Delta y\big)
+ \Delta\lambda\bigg)\exp{\big(-\frac{3}{2}\tau\big)}\cos{\big(n\tau\big)}+ \nonumber \\
& & + \frac{3}{8n}\bigg(-\frac{2(4n^{2}+9)(1+2w_{m})(1-3w_{m})}{4n^{2}+9(1+2w_{m})^{2}} 
\big(\Delta x -2 \Delta y\big) 
+\nonumber \\ & & \hspace{1.25cm}
+ \delta\Delta x+ 4\Delta\lambda\bigg)
\exp{\big(-\frac{3}{2}\tau\big)}\sin{\big(n\tau\big)}\,.
\end{eqnarray}
\end{subequations}
where $\Delta x = x^{(i)}$, $\Delta y = y^{(i)}-1$, and $\Delta\lambda=\lambda^{(i)}+2$ are the 
initial conditions.
Again, up to linear terms in initial conditions, we obtain linearized solution for 
the barotropic matter density fraction 
\begin{equation}
\label{eq:Omega_m2}
\Omega_{m} \approx (\Delta x - 2 \Delta y)\exp{\big(-3(1+w_{m})\tau\big)} = \Omega_{m,i}
\exp{\big(-3(1+w_{m})\tau\big)}.
\end{equation}

Using these solutions and the acceleration equation \eqref{eq:acc:3}, one is able to obtain the 
Hubble function in the vicinity of the critical point of the focus type. Up to linear terms in 
initial conditions we have
\begin{equation}
\label{eq:H2T_2}
\left(\frac{H(a)}{H(a_{0})}\right)^{2}  \approx  \Omega_{\Lambda,0} +  
\Omega_{M,0}\left(\frac{a}{a_{0}}\right)^{-3(1+w_{m})} 
+\left(\frac{a}{a_{0}}\right)^{-3/2}
\Bigg(
\Omega_{cos,0}\,\cos{\left(n\,\ln{\bigg(\frac{a}{a_{0}}\bigg)}\right)} +
\Omega_{sin,0}\,\sin{\left(n\,\ln{\bigg(\frac{a}{a_{0}}\bigg)}\right)}
\Bigg)
\end{equation}
where
\begin{equation}
\Omega_{\Lambda,0} = 1-\Omega_{M,0}-\Omega_{cos,0}
\end{equation}
and
\begin{subequations}
\begin{align}
\Omega_{M,0} = & \bigg(1-\frac{4(4n^{2}+9)(1-3w_{m})(4+3w_{m})}{3\delta(4n^{2}+9(1+2w_{m})^{2})}
\bigg)\Omega_{m,0}\,,\\
\Omega_{cos,0} = & 
\frac{1}{3\delta}\bigg(\frac{4(4n^{2}+9)(1-3w_{m})(4+3w_{m})}{4n^{2}+9(1+2w_{m})^{2}}
\Omega_{m,i} 
-3\delta\,\Delta x + 8 \Delta\lambda\bigg)
\left(\frac{a_{0}}{a^{(i)}}\right)^{-3/2}\cos{\left(n\,\ln{\bigg(\frac{a_{0}}{a^{(i)}}\bigg)}
\right)} + \nonumber \\
 + &\frac{1}{6\delta n}\bigg(\frac{2(4n^{2}+9)(1-3w_{m})(4n^{2}-15(1+2w_{m}))}
{4n^{2}+9(1+2w_{m})^{2}}
\Omega_{m,i} 
+ \nonumber \\ & \hspace{1cm}
+ 15\delta\,\Delta x + 4(4n^{2}+15) \Delta\lambda
\bigg)\left(\frac{a_{0}}{a^{(i)}}\right)^{-3/2}\sin{\left(n\,\ln{\bigg(\frac{a_{0}}{a^{(i)}}
\bigg)}\right)}\,,\\
\Omega_{sin,0} = &\frac{1}{6\delta n}\bigg(\frac{2(4n^{2}+9)(1-3w_{m})(4n^{2}-15(1+2w_{m}))}
{4n^{2}+9(1+2w_{m})^{2}}
\Omega_{m,i} 
+ \nonumber \\ & \hspace{1cm}
+ 15\delta\,\Delta x + 4(4n^{2}+15) \Delta\lambda
\bigg)\left(\frac{a_{0}}{a^{(i)}}\right)^{-3/2}\cos{\left(n\,\ln{\bigg(\frac{a_{0}}{a^{(i)}}
\bigg)}\right)} - \nonumber \\
-&\frac{1}{3\delta}\bigg(\frac{4(4n^{2}+9)(1-3w_{m})(4+3w_{m})}{4n^{2}+9(1+2w_{m})^{2}}
\Omega_{m,i} 
- \nonumber \\ &\hspace{1cm}
-3\delta\,\Delta x + 8 \Delta\lambda\bigg)
\left(\frac{a_{0}}{a^{(i)}}\right)^{-3/2}\sin{\left(n\,\ln{\bigg(\frac{a_{0}}{a^{(i)}}\bigg)}
\right)}\,. 
\end{align}
\end{subequations}

In this section we obtained two forms of Hubble functions for two different types 
of behavior in the vicinity of the critical point representing the de Sitter state. 

\subsection{Dust matter}

For dust matter and special initial conditions,
\begin{equation}
\Delta x = \frac{4}{\delta}\Omega_{m,i}\,,\qquad \Delta \lambda = -\frac{1}{2}\Omega_{m,i}
\label{eq:init}
\end{equation}
where from \eqref{eq:Omega_m1} and \eqref{eq:Omega_m2} we have $\Omega_{m,i}= \Delta x -2 \Delta y$, and both forms of the Hubble functions \eqref{eq:H2T_1} and \eqref{eq:H2T_2} take
the same form. Note that these initial conditions
do not define a single point in three-dimensional phase space but a set of initial
conditions. In the case of the monotonic approach to the de Sitter
state \eqref{eq:H2T_1}, we have $\Omega_{n,0}=0$ and $\Omega_{3n,0}=0$, and in
the case of the oscillatory approach \eqref{eq:H2T_2}, we have
$\Omega_{cos,0}=0$ and $\Omega_{sin,0}=0$ and the resulting form of the
Hubble function is
\begin{equation}
\left(\frac{H(a)}{H(a_{0})}\right)^{2} \approx 1 - \Omega_{M,0} + \Omega_{M,0}\left(\frac{a}{a_{0}}\right)^{-3}
\end{equation}
where
\begin{equation}
\Omega_{M,0} = \bigg(1-\frac{16}{3\delta}\bigg)\Omega_{m,0}\,.
\label{eq:OmegaM}
\end{equation}
Note that when the parameter $\delta$ in Eq.~\eqref{eq:OmegaM} is negative $\delta<0$, the total matter
density observed in the universe $\Omega_{M,0}$ is larger than the matter density included in
the model by hand $\Omega_{m,0}$.

If we assume that in the model we include only the baryonic matter
$\Omega_{m,0}=\Omega_{bm,0}$, then from the recent astronomical observations of
the Planck satellite \cite{Ade:2013lta}, we have that the present total matter
density parameter is $\Omega_{M,0}\approx0.315$ and the baryonic matter
density parameter is $\Omega_{bm,0}\approx0.049$. This gives us an opportunity to
find the value of the $\delta$ parameter of the model, which gives us information
about the second derivative of the potential function at the critical point.
After a little algebra, one finds
\begin{equation}
\delta=\frac{16}{3\left(1-\frac{\Omega_{M,0}}{\Omega_{bm,0}}\right)} \approx -0.9825\,.
\end{equation}
Next, from the definition of the $\delta$ parameter, one can directly calculate
the value of the second derivative of the scalar field potential at the
critical point
\begin{equation}
\Gamma\big|_{*} = \frac{V''(\phi) V(\phi)}{V'(\phi)^{2}}\Big|_{*} \approx 0.5921\,,
\end{equation}
which indicates that at the critical point, the second derivative of the potential function is positive $V''(\phi)\big|_{*}>0$.

Then from the assumed initial conditions \eqref{eq:init}, one gets the present values of the phase space variables,
\begin{equation}
\label{eq:41}
x(a_{0})=\frac{\dot{\phi}}{H\phi}\Big|_{0} \approx -0.1995\,,\qquad \lambda(a_{0}) = -\phi\frac{V'(\phi)}{V(\phi)}\Big|_{0} \approx -2.0245\,.
\end{equation}

From the action integral \eqref{eq:action} one can define the effective gravitational coupling of Brans-Dicke
theory as
\begin{equation}
G_{\text{eff}}=\frac{1}{\phi}\,.
\label{eq:Geff}
\end{equation}
However, the effective gravitational coupling as determined in Cavendish-type experiments is
slightly different. The spherically symmetric solution in the Brans-Dicke theory yields
\cite{EspositoFarese:2000ij, Damour:1992we, Damour:1995kt}
\begin{equation}
G_{\text{eff}}=\frac{1}{\phi}\frac{4+2\om}{3+2\om}\,.
\label{eq:Gsol}
\end{equation}
We must bear in mind that this quantity is defined in the context of the parametrized
post-Newtonian (PPN) formalism \cite{Will:book} for spherically symmetric solutions suitable
for solar system tests, and not for cosmological ones \cite{Faraoni:book}.

The use of \eqref{eq:Geff} or \eqref{eq:Gsol} gives the variation of the effective gravitational
coupling in the Brans-Dicke theory as
\begin{equation}
\frac{\dot{G}_{\text{eff}}}{G_{\text{eff}}} = - \frac{\dot{\phi}}{\phi}\,,
\end{equation}
and the time variation is due to the cosmological evolution of the scalar field $\phi$
\cite{Faraoni:book}. Now, from the definition of the phase space variable $x$, we obtain that its
present value is given by
\begin{equation}
x(a_{0})=\frac{\dot{\phi}}{H\phi}\Big|_{0} = - \frac{\dot{G}}{H G}\Big|_{0}\,,
\end{equation}
where in $G$ we omitted the subscript for simplicity.

Now from \eqref{eq:41}, we obtain
\begin{equation}
\frac{\dot{G}}{H G}\Big|_{0} \approx 0.1995
\end{equation}
which, taking the present age of the Universe $t_{0}=13.817\times10^{9} {\rm yr}$, gives
\begin{equation}
\frac{\dot{G}}{G}\Big|_{0} \approx 1.44\times10^{-11}\frac{1}{\rm yr}\,,
\end{equation}
and is in good agreement with other observational constraints on the variability of the gravitational coupling constant \cite{Uzan:2010pm}.

One can also calculate the mass of the Brans-Dicke scalar field. In the Jordan frame it is given by \cite{Faraoni:2009km}
\begin{equation}
m^{2} = \frac{2}{3+2\om}\big(\phi V''(\phi)-V'(\phi)\big),
\end{equation}
which in the previously introduced phase space variables becomes
\begin{equation}
m^{2} = \frac{6}{3+2\om}H^{2}y^{2}\lambda\big(1+\lambda\Gamma(\lambda)\big)\,.
\end{equation}
Direct calculation of the asymptotic value of the scalar field mass at the critical point gives
\begin{equation}
m^{2}\big|_{*} = -\frac{9}{4} H_{*}^{2}\frac{\delta}{3+2\om}\,
\end{equation}
where $H_{*}^{2} \approx H_{0}^{2}(1-\Omega_{M,0})$ is the asymptotic value of the Hubble 
function and $H_{0}$ its present value. Inserting the estimated value of the $\delta$ parameter 
and $H_{0} \approx 1.5 \times10^{-33}{\rm eV}$, one obtains
\begin{equation}
m\big|_{*} \approx 1.84 \frac{10^{-33}}{\sqrt{3+2\om}}\,{\rm eV}\,,
\end{equation}
which is an asymptotic value of the mass of the Brans-Dicke scalar field at the de Sitter state. 
Such ultralight scalar particles are usually postulated to explain the dynamic of galaxies \cite{Lora:2011yc} and are treated as bosonic cold dark matter candidates \cite{RodriguezMontoya:2010zza,Park:2012ru}. To obtain the mass of the scalar particle of order $10^{-22}\,{\rm eV}$, one needs
\begin{equation}
\om \approx -\frac{3}{2}+10^{-22}\,,
\end{equation}
which is very close to the conformal coupling value \cite{Faraoni:book}.

The Cassini spacecraft mission in the parametrized post-Newtonian (PPN) formalism gave
the most stringent experimental limit $\om>40000$ on the value of the Brans-Dicke
parameter \cite{Bertotti:2003rm}. This was obtained in the solar system test for
spherically symmetric solutions. On the other hand, in cosmology when homogeneity and
isotropy of the space are assumed, the cosmography plays the role of the PPN formalism.
The model survives the cosmographic test when it predicts the correct values of the
cosmographic parameters \cite{Capozziello:book}. In the model under
consideration the agreement with the standard cosmological model --- the $\Lambda$CDM model --- was
obtained at the very beginning.

The huge inconsistency of the obtained values of the Brans-Dicke parameter $\om$ can be approached twofold. First, from the methodological viewpoint, in cosmology, within a given theory of gravity one builds models taking some assumptions about geometry of space and then constrains these models with observational data. Models have limitations of applicability, and extrapolations of the results beyond their performance range leads to controversies, like the one discussed here. Second, this inconsistency can be understood through screening mechanisms \cite{Khoury:2010xi, Khoury:2013yya}, which rely on the suppression of the deviations from standard gravity by the high density of the local settings. Two of them, the chameleon mechanism \cite{Khoury:2003aq, Khoury:2003rn, Gubser:2004uf, Brax:2004qh, Mota:2006ed, Mota:2006fz} and the symmetron mechanism \cite{Hinterbichler:2010es, Olive:2007aj, Pietroni:2005pv, Hinterbichler:2011ca, Brax:2011pk}, are closely related. The chameleon mechanism leads to modifications in the effective mass of the scalar field which depends on the local matter density. In regions of low mass density, the scalar field is light, while in regions of high density, it acquires a large mass, making its effects unobservable. The symmetron mechanism involves the vacuum expectation value of a scalar field that depends on the local mass density. In regions of low mass density, the vacuum expectation value of a scalar field becomes large, while in regions of high mass density, it becomes small. In the symmetron mechanism the scalar couples with gravitational strength in regions of low density; it and decouples and is screened in regions with high mass density. The last of the screening mechanisms is the Vainshtein mechanism \cite{Vainshtein:1972sx, ArkaniHamed:2002sp, Deffayet:2001uk}, which in the vicinity of massive sources leads to the large derivative coupling of a scalar field. The nonlinear interactions crank up the kinetic term of perturbations, leading to weaker interactions with matter. 

Our considerations, taking no account of the contentious question concerning the Einstein frame or the Jordan frame \cite{Faraoni:1999hp}, are similar to the chameleon mechanism, where matter couples directly with the scalar field.

Now we show that the $\Lambda$CDM model can also emerge without assumption
about specific initial conditions but rather as a specific assumption about
the model parameters. For dust matter $w_{m}=0$, Eqs.~\eqref{eq:Omega_1} reduce to
\begin{subequations}
\begin{align}
\Omega_{M,0} = & \bigg(1-\frac{16}{3\delta}\bigg)\Omega_{m,0}\,,\\
\Omega_{n,0} = & \frac{n+1}{3\delta(2n-3)} \bigg(4n\,\Omega_{m,i}-3\delta\Delta x + 8(n-3)\,
\Delta
\lambda\bigg)\left(\frac{a_{0}}{a^{(i)}}\right)^{-n}\,,\\
\Omega_{3n,0} = & \frac{n-4}{3\delta(2n-3)}\bigg(-4(n-3)\,\Omega_{m,i} - 3\delta\Delta x - 8n\, 
\Delta\lambda\bigg)\left(\frac{a_{0}}{a^{(i)}}\right)^{-3+n}\,.
\end{align}
\end{subequations}
Expanding these quantities for $|n| \ll 1$ and assuming that $n^{2}\approx0$ and $n\Delta x=n\Delta y=n\Delta\lambda\approx0$, one obtains
\begin{subequations}
\begin{align}
\Omega_{n,0} \approx & \frac{1}{3\delta}\bigg(\delta\Delta x + 8 \Delta\lambda\bigg)\,,\\
\Omega_{3n,0} \approx & \frac{4}{9\delta}\bigg(12 \Omega_{m,i} -3\delta\Delta x\bigg) 
\left(\frac{a_{0}}{a^{(i)}}\right)^{-3}\,,
\end{align}
\end{subequations}
and the Hubble function \eqref{eq:H2T_1} is
\begin{equation}
\left(\frac{H(a)}{H(a_{0})}\right)^{2} \approx 1 - \bigg(\Omega_{m,0} - \frac{4}{3}\Delta x 
\left(\frac{a_{0}}{a^{(i)}}\right)^{-3}\bigg) + 
\bigg(\Omega_{m,0} - \frac{4}{3}\Delta x \left(\frac{a_{0}}{a^{(i)}}\right)^{-3}\bigg) 
\left(\frac{a}{a_{0}}\right)^{-3}\,.
\end{equation}
Note that in the general case of the monotonic approach to the de Sitter state, we have
\begin{equation}
\frac{\delta}{3+2\om}=\frac{4}{9}n(n-3)\,
\end{equation}
and the assumption that $|n|\ll1$ should be interpreted as the smallness of the ratio,
\begin{equation}
\frac{\delta}{3+2\om}\approx-\frac{4}{3}n\,.
\end{equation}

From \eqref{eq:linT_1a} we have 
$x(a)\approx \Delta x \left(\frac{a}{a^{(i)}}\right)^{-3}$, which gives 
$x(a_{0})\approx \Delta x\left(\frac{a_{0}}{a^{(i)}}\right)^{-3}$. From the definition of the 
variable $x$, we obtain that at the present epoch,
\begin{equation}
x(a_{0}) = \frac{\dot{\phi}}{H \phi}\Big|_{0}.
\end{equation}
If we assume that in the model we include only the baryonic matter $\Omega_{m}=\Omega_{bm}$ and 
note that in the Brans-Dicke theory the field $\phi$ can be identified as the inverse of the 
effective gravitational coupling \eqref{eq:Geff}
which is now the function of the spacetime location, we obtain the following form of the Hubble 
function,
\begin{equation}
\left(\frac{H(a)}{H(a_{0})}\right)^{2} \approx 1 - \bigg(\Omega_{bm,0}+\Omega_{dm,0}\bigg) + 
\bigg(\Omega_{bm,0} + \Omega_{dm,0}\bigg) \left(\frac{a}{a_{0}}\right)^{-3}\,,
\end{equation}
where the present dark matter density parameter is 
\begin{equation}
\Omega_{dm,0} = \frac{4}{3}\frac{\dot{G}}{H G}\Big|_{0}\,.
\end{equation}
From the Planck satellite data \cite{Ade:2013lta}, we have the present
value of the density parameter of the dark matter $\Omega_{dm,0}\approx0.266$,
and we obtain
\begin{equation}
\frac{\dot{G}}{H G}\Big|_{0}\approx0.1995\,.
\end{equation}
Two different assumptions lead to very similar results.

\subsection{Low-energy string theory limit}

The Lagrangian density of the low-energy limit of the bosonic string theory
\cite{Green:book,Fradkin:1985ys,Tseytlin:1988rr} can be presented in the following form,
\begin{equation}
\mathcal{L} = e^{-2\Phi}\big(R + 4\nabla^{\alpha}\Phi\,\nabla_{\alpha}\Phi - \Lambda\big)
\end{equation}
where $\Phi$ is the dilaton field. Making the substitution $\phi=e^{-2\Phi}$, one obtains the 
Brans-Dicke theory with $\om=-1$ and $V(\phi)=\Lambda\phi$. Neglecting the matter, the two 
theories are identical, but they differ in their couplings of the scalar field to the other 
matter \cite{Kolitch:1994qa}.

The constant value of the Brans-Dicke parameter $\om=-1$ leads to
\begin{equation}
\delta= \frac{4}{9}n(n-3)\,,
\end{equation}
and for dust matter $w_{m}=0$, Eqs.~\eqref{eq:Omega_1} reduce to
\begin{subequations}
\begin{align}
\Omega_{M,0} = & \bigg(1-\frac{12}{n(n-3)}\bigg)\Omega_{m,0}\,,\\
\Omega_{n,0} = & \frac{n+1}{n(n-3)(2n-3)} \bigg(3n\,\Omega_{m,i}-n(n-3)\Delta x + 6(n-3)\,\Delta
\lambda\bigg)\left(\frac{a_{0}}{a^{(i)}}\right)^{-n}\,,\\
\Omega_{3n,0} = & \frac{n-4}{n(n-3)(2n-3)}\bigg(-3(n-3)\,\Omega_{m,i} - n(n-3)\Delta x - 6n\, 
\Delta\lambda\bigg)\left(\frac{a_{0}}{a^{(i)}}\right)^{-3+n}\,.
\end{align}
\end{subequations}
In this case, we also expand these quantities for $|n|\ll 1$, which indicates
that $|\delta|\ll 1$, and we are very close to the quadratic potential
function.
The resulting Hubble function up to linear terms in the initial condition is
\begin{equation}
\left(\frac{H(a)}{H(a_{0})}\right)^{2} \approx  1 - \frac{2}{3}\bigg(\Omega_{m,0} - 
2 x(a_{0})+4\Delta\lambda\bigg) + 
\frac{2}{3}\bigg(\Omega_{m,0} - 2x(a_{0})+4\Delta\lambda\bigg)
\left(\frac{a}{a_{0}}\right)^{-3}
+ \bigg(2\Delta\lambda - 4 \Omega_{m,0}\left(\frac{a}{a_{0}}\right)^{-3}\bigg)
\ln{\left(\frac{a}{a_{0}}\right)}\,.
\end{equation}
One notices that in the case of the low-energy string theory limit, we obtain a
different form of the Hubble function, applying the same type of expansion,
but the difference lies in the direct connection of the $n$ parameter with the
shape of the potential function at the critical point.

The resulting Hubble function differs by the presence of the term
proportional to the natural logarithm of the scale factor. At the present epoch
this term vanishes, but in the past it could play a critical role.

\section{Conclusions}

Investigating the emergence of the standard cosmological model ($\Lambda$CDM) from the Brans-Dicke 
theory of gravity has shown the significant problem in modern cosmology, namely, the problem 
of initial conditions. In cosmology we have no knowledge of the initial conditions for the evolution of the Universe, which is why we try not to be bound to a very specific initial condition, but to have many possible ones that evolve to the known Universe. To address this problem, we have studied all the possible evolutions for all admissible initial conditions. Then we have tested them against the astronomical data to choose empirically which evolutionary scenario is realized for our Universe.

We have shown that the dynamical systems methods are especially useful when we are interested in an initial
condition problem, so we study all the evolutionary paths for all admissible initial conditions. However, cosmology is not just pure mathematics, because we can constrain possible evolutions
(represented by the trajectories in the phase space) by the observations, that is we need to find the initial conditions. Cosmology is the science of the evolution of the universe as well as the initial conditions that give rise to this evolution.

In this paper we have shown how the standard cosmological model ($\Lambda$CDM) emerges from the Brans-Dicke theory. There are two ways of approaching the $\Lambda$CDM model: a monotonic one and an oscillating one. Which specific case takes place crucially depends on the initial conditions for the Universe as well as the model parameters. While the leading terms in both cases represent the $\Lambda$CDM-type evolution, there are additional terms that describe the asymptotic state of the system (a node or a focus type critical point in the phase space). The density parameters for the dark energy as well as for the dark matter are emergent parameters from the Brans-Dicke cosmology. The values of these parameters are fragile due to the value and the sign of the second derivative of the scalar field potential at the critical point.

Finally, from the recent Planck satellite data, we have obtained the constraint on the variability of the effective gravitational coupling constant, $\frac{\dot{G}}{G}\Big|_{0} \approx 1.44\times10^{-11}\frac{1}{\rm yr}\,$, as well as the lower limit of the mass of the Brans-Dicke scalar field at the de Sitter state $m\big|_{*} \approx 1.84 \frac{10^{-33}}{\sqrt{3+2\om}}\,{\rm eV}\,$. Our results are do not depend on the specific form of the potential function for the Brans-Dicke scalar field.

\begin{acknowledgments}
 We are very grateful to the organizers of the 49th Winter School of Theoretical Physics,
``Cosmology and non-equilibrium statistical mechanics'', L{\c a}dek-Zdr{\'o}j, Poland, February
10-16, 2013, especially to Professor Andrzej Borowiec for the invitation and opportunity to present 
part of this work. 
We thank Konrad Marosek for discussion and comments.

The work of O.H. was supported by the Polish Ministry of Science and Higher
Education project ``Iuventus Plus'' (Contract No.~0131/H03/2010/70) and by the
National Science Centre through the postdoctoral internship award (Decision
No.~DEC-2012/04/S/ST9/00020).
\end{acknowledgments}

\bibliographystyle{apsrev4-1}
\bibliography{../bd_theory,../darkenergy,../quintessence,../quartessence,../astro,../dynamics,../standard,../inflation,../sm_nmc,../singularities,../strings}

\end{document}